# Real time optical observation and control of atomically thin transition metal dichalcogenide synthesis

Hamid Reza Rasouli[1], Naveed Mehmood[1], Onur Çakıroğlu[2], T. Serkan Kasırga[1,2]*

[1] UNAM – Institute of Materials Science and Nanotechnology, Bilkent University, Ankara 06800, Turkey

[2] Department of Physics, Bilkent University, Ankara 06800, Turkey

* Corresponding Author: kasirga@unam.bilkent.edu.tr

**Abstract:**

**Understanding the mechanisms involved in chemical vapour deposition (CVD) synthesis of atomically thin transition metal dichalcogenides (TMDCs) requires the precise control of numerous growth parameters. All the proposed mechanisms and their relation to the growth conditions are inferred from characterising intermediate formations obtained by stopping the growth blindly. To fully understand the reaction routes that lead to the monolayer formation, real time observation and control of the growth are needed. Here, we demonstrate how a custom-made CVD chamber that allows real time optical monitoring can be employed to study the reaction routes that are critical to the production of the desired layered thin crystals in salt assisted TMDC synthesis. Our real time observations reveal the reaction between the salt and the metallic precursor to form intermediate compounds which lead to the layered crystal formation. We identified that both the vapour-solid-solid and vapour-liquid-solid growth routes are in an interplay. Furthermore, we demonstrate the role $H_2$ plays in the salt-assisted $WSe_2$ synthesis. Finally, we guided the crystal formation by directing the liquid intermediate compound through pre-patterned channels. The methods presented in this article can be extended to other materials that can be synthesized via CVD.**

**Main Text:**

Chemical vapour deposition (CVD) synthesis of two-dimensional (2D) transition metal dichalcogenides (TMDCs) involves deposition of gaseous precursors on to a substrate to facilitate the crystallization in the desired crystal structure[1,2,3,4,5]. In a typical CVD synthesis, a transition metal containing precursor is placed in a tube furnace with a chalcogen precursor and a target substrate. $Ar/H_2$ mixture carries the vaporised chalcogen precursor and the metal compounds to form atomically thin layers on the target substrate. Salts are also added to the conventionally used metal oxide precursors to form more volatile intermediate compounds[6,7,8]. This increases the monolayer formation rate and allows the synthesis of otherwise difficult to synthesize 2D TMDCs[9]. The setup described above has been used to produce atomically thin TMDC crystals in various morphologies. However, optimization of the growth parameters requires blind trial and errors, and even the optimized recipes offer limited control in terms of number of layers, crystal phase and morphology.

There are two growth modes in CVD synthesis of TMDCs. (1) Vapour-Solid-Solid (VSS): Vaporized precursors are adsorbed on the substrate and form crystals via surface diffusion and bond formation at an elevated temperature[10], and (2) Vapour-Liquid-Solid (VLS): Supersaturated liquid droplets containing the constituent elements form the crystals[11]. **Figure 1a** depicts these growth modes. Despite many studies on the CVD growth mechanisms of few layer TMDCs, it is unclear which growth mode prevails under different growth conditions. The greatest challenge in understanding the on-going processes during the growth is the inaccessibility of the tube furnace for real time observations. To analyse the intermediate products leading to the desired crystal growth, these products must be caught blindly by shutting off the furnace and quenching the synthesis by a rapid cool down. Although there are reports on real time visual observation of graphene[12–14], $Y_2BaCuO_5$ and[15] vanadium dioxide nanocrystal[16] synthesis, due to the complexity of the growth process no such observations have been made for TMDCs.

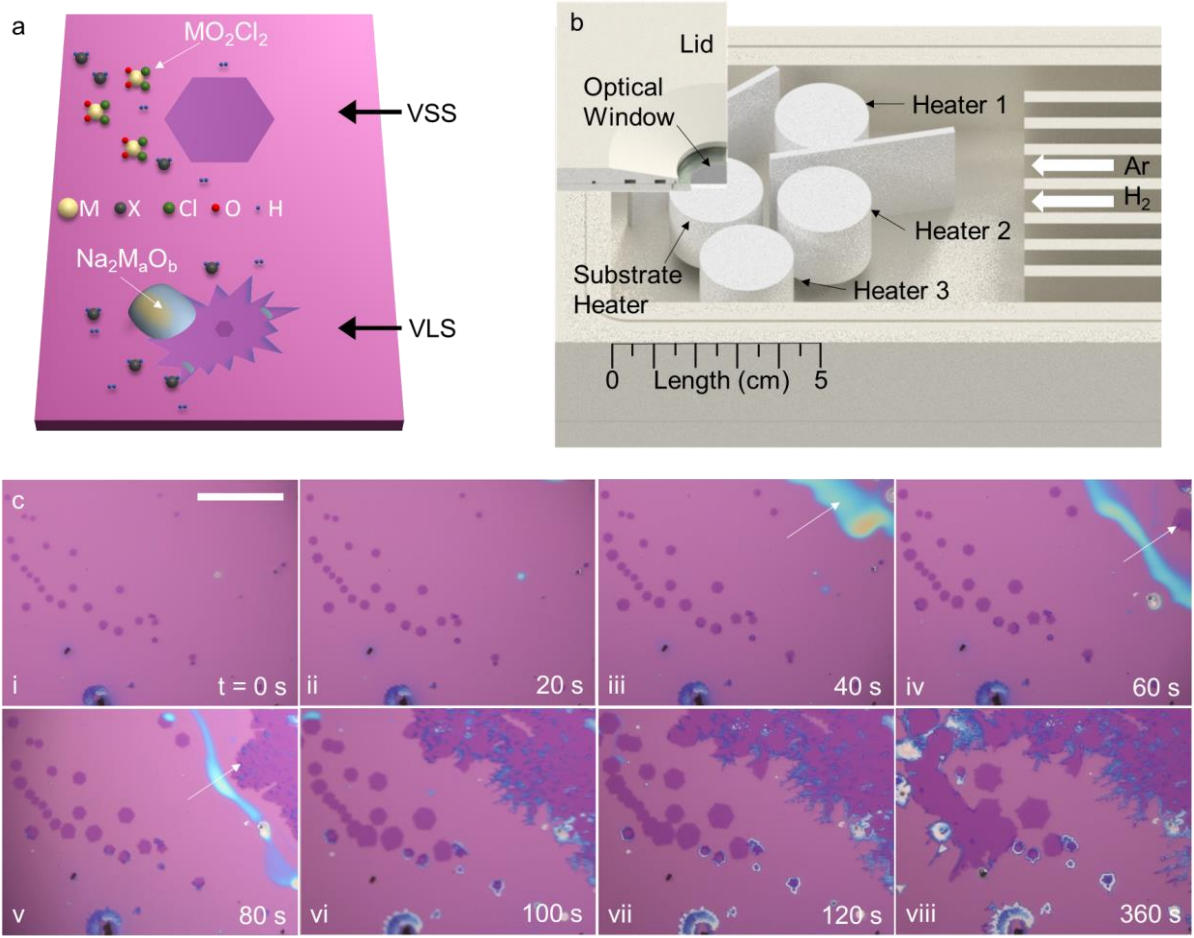

**Figure 1 | Schematic of the CVD growth modes, custom-made CVD chamber, and an exemplary growth. a,** Depictions of VSS and VLS growth modes for NaCl assisted TMDC synthesis. In VSS mode, adsorbed vapour forms the crystals while in VLS mode, a liquid precursor forms the crystal via a chemical reaction. M and X denote the metal and the chalcogen precursors, respectively. **b,** Schematic of the custom-made CVD chamber depicts the configuration of the heaters, relevant lengths and the quarter cross-section of the upper lid. Ar and $H_2$ flow is indicated by the white arrows. Heater 1 is separated from the rest of the reactors to prevent heating of the chalcogen precursor by radiation from the precursor heaters. **c**, Real time optical micrographs captured during a synthesis is given in **i-viii**. The substrate temperature is at 790 °C and the series of pictures show the evolution of atomically thin $WSe_2$ crystals. Hexagonal monolayers grow larger as time goes on with a rate of ~0.2 μm/s. White arrow on **iii** attracts attention to the liquid that promotes the synthesis of the monolayers. As time goes on the amount of liquid diminishes and irregularly shaped $WSe_2$ crystals form. After 6 minutes (**viii**) from the beginning of the observation, no liquid is left, and the lateral growth of the crystals stops. Scale bar is 100 μm.

To explore the synthesis of atomically thin TMDCs, we build a custom-made CVD chamber that allows real time optical observation and control of multi-precursor crystal growth. **Figure**

**1b** shows a schematic of the chamber. Our investigations of the CVD growth mechanisms of atomically thin TMDCs rest on the ability to control four separate alumina heaters while monitoring the growth substrate under an optical microscope. This ability allows the chamber to be used as a multi-zone chamber. One heater dedicated for the growth substrate is directly located under a 0.5 mm thick sapphire window for optical observations. A 40x ultra-long working distance (4.4 mm) objective is employed to have a high resolving power while maintaining a large enough hot zone above the substrate. The other heaters are dedicated for the growth precursors and their separation to the substrate heater can be adjusted within the chamber (see Supplementary Information for details).

We focus on salt assisted synthesis of $WSe_2$ monolayers on oxidised Si chip as a demonstration of the versatility of our chamber. **Figure 1c** shows a series of optical images taken during the synthesis at 790 °C (see Supplementary Movie 1). $WSe_2$ crystal formation via VSS and VLS modes can be observed in real time. Fine mesh grains of $WO_3$ and NaCl are placed on heater 2 and Se on heater 1. The substrate heater and heater 2 temperatures are increased simultaneously. When they reach 600 °C, heater 1 is brought to 300 °C, above the melting point of Se. At the same time $H_2$ is introduced to aid the growth of the monolayers. Atomic force microscope (AFM) image, Raman and photoluminescence (PL) maps taken from typical crystals show that they are high quality $WSe_2$ monolayers (Supplementary Information **Figure S3** and **S4**).

To understand the possible crystal formation routes, we placed a 250 µm³ large grain of NaCl surrounded by smaller grains of $WO_3$ on a $SiO_2$/Si chip on the substrate heater and observed the dynamics of the intermediate compound formation before $WSe_2$ growth. Optical images captured during the heat up show the intermediate stages of the reaction between $WO_3$ and NaCl (**Figure 2a-f**). Firstly, we observe a turquoise liquid formation in the vicinity of the $WO_3$ particles as low as 600 °C. This is consistent with the previously reported thermogravimetry and differential scanning calorimetry measurements on NaCl-$WO_3$ mixtures[9]. At this early stage, when we stop the synthesis and perform scanning electron microscopy (SEM) imaging and energy dispersive X-ray spectroscopy (EDX) mapping, Na and Cl on the $WO_3$ grains are detected (see **Figure S5** in Supplementary Information). This clearly indicates that NaCl sublimates and condenses on $WO_3$. At higher substrate temperatures, more liquid forms and its colour changes from turquoise to beige. Considering that both NaCl and $WO_3$ has melting points above 800 °C, the liquid formation is a product of the reaction between the two. Especially above 730 °C, as time goes on, no solid $WO_3$ grains remain as a result of reaction with NaCl (see Supplementary Movie 2). As reported previously in the literature[11] and as illustrated later in the text, this reaction has a significant role in both VLS and VSS crystal synthesis.

Once the liquid is fully formed, we cool down the chamber and immediately perform X-ray photoelectron spectroscopy (XPS) on the solidified liquid to determine its chemical composition. XPS surveys on the solidified liquid show that it is composed of Na, O, W but not Cl. High resolution XPS spectra of Na $1s$, O $1s$ and W $4f$ binding energies exactly match with the values reported for sodium tungstate, $Na_2WO_4$ (**Figure 2g**)[17,18]. The measured Raman spectrum is in good agreement with the reported spectrum for $Na_2WO_4$ in the literature (**Figure 2h**)[19,20]. When we consider the chemical reaction between NaCl and $WO_3$ the second product of the reaction can be $WO_2Cl_2$ [9,11]:

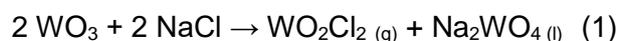

$$2\ WO_3 + 2\ NaCl \rightarrow WO_2Cl_{2\ (g)} + Na_2WO_{4\ (l)} \quad (1)$$

At above 600 °C, the temperature where we observe the liquid formation, $WO_2Cl_2$ is in gaseous phase. As the reaction between vapour NaCl and $WO_3$ begins, $WO_2Cl_2$ forms and leaves the chamber with the carrier gas. This explains why we don't observe any Cl both in XPS and EDX analysis of the later stage molten product. Also, as discussed later in the text, by controlling the presence of $WO_2Cl_2$ in the chamber, we can lead the growth to follow either VLS or VSS mode.

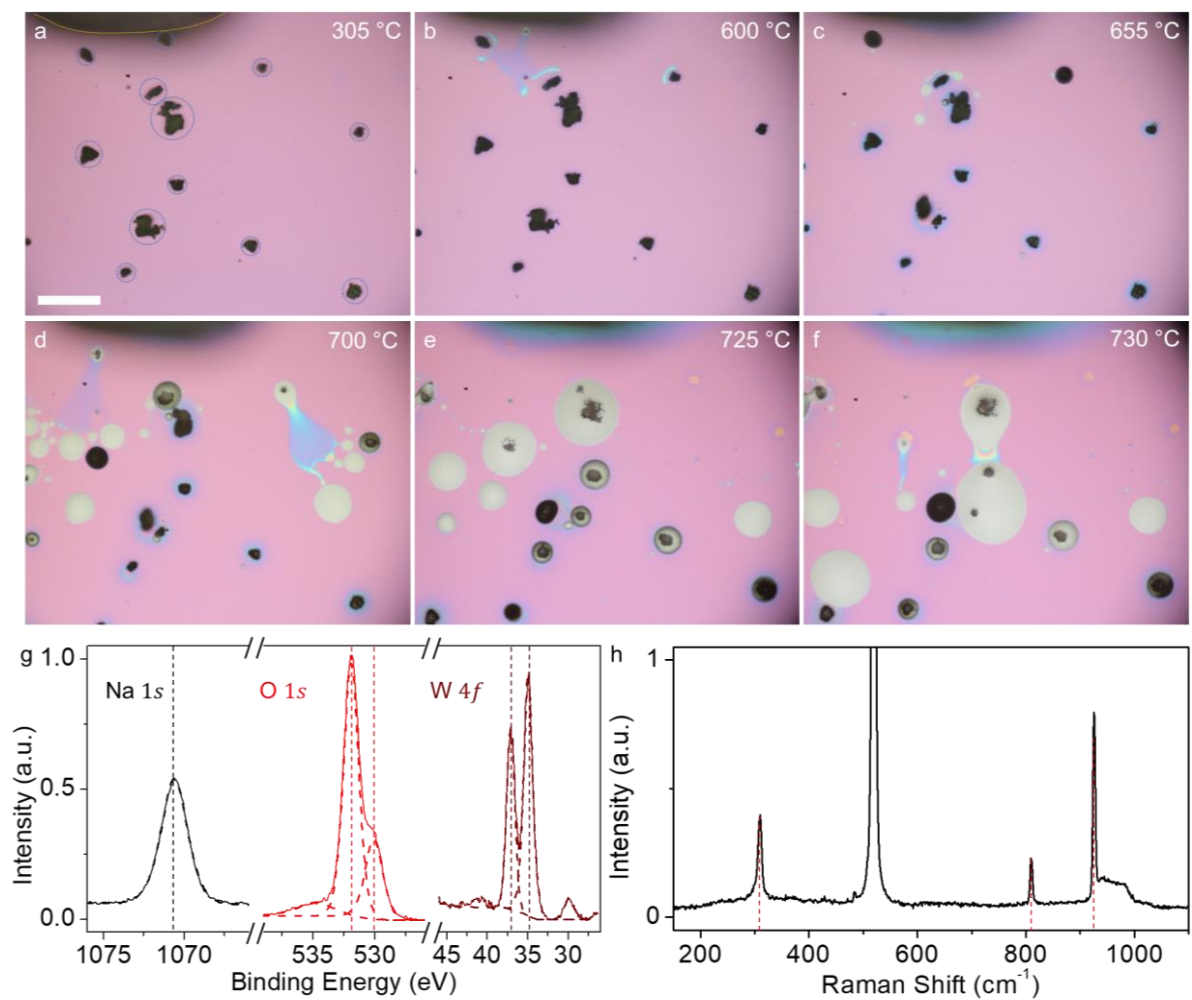

**Figure 2 | Formation and characterization of the intermediate compound. a-f,** Optical micrographs taken during the synthesis at various temperatures. Yellow dashed line at the top in **a** indicates the edge of the salt particle and grains encircled by blue dashed lines are $WO_3$ particles. As the temperature increases first a turquoise liquid forms around the $WO_3$ grains. Then, as time goes on the liquid droplet enlarges by consuming the $WO_3$ grains. Scale bar is 100 µm. **g**, XPS survey shows that the liquid is composed of Na, O and W. Na $1s$ survey can be fitted with a signle peak at 1070.6 eV that can be attributed to Na in $Na_2WO_4$. O $1s$ binding energy can be fitted with two peaks, one at 532.9 eV and another at 530.2 eV that can be attributed to the substrate and to $Na_2WO_4$, respectively. W $4f^{5/2}$ and $4f^{7/2}$ peaks are located at 37.3 and 34.9 eV corresponds to the previously reported values[17,18] for W $4f$ of $Na_2WO_4$ in the literature. The peak at 30.1 eV can be attributed to Na $2p$. **h**, Raman spectrum taken from a solidified intermediate compound. Peaks marked with red dashed lines correspond to the Raman modes of $Na_2WO_4$.

With the unique abilities we have with our custom-made CVD chamber, we investigated the possible monolayer $WSe_2$ formation routes from both liquid and gaseous intermediate compounds. First, we prepared a mixture of $NaCl:WO_3$ in 1:2 weight ratio to study how molten $Na_2WO_4$ forms $WSe_2$ monolayers. 1:1 molar ratio of $NaCl:WO_3$ (based on Reaction 1) corresponds to 1:4 weight ratio, yet the salt rich mixture results in liquid $Na_2WO_4$ with minimum solid content as some salt sublimates during the heat up. A grain of the mixture is placed on an oxidised silicon chip and then heated to 750 °C under Ar environment on the substrate heater to form liquid $Na_2WO_4$. After the liquid formation, we introduced Se. However, this didn't result in $WSe_2$ synthesis and the liquid remained unchanged as the time went on. Although we tried introducing Se vapour at various substrate temperatures ranging from 700 to 900 °C, we didn't observe $WSe_2$ formation. Thus, we deduce that $H_2$ plays an essential role in the $WSe_2$ formation.

Although there are numerous reports[21–26] on the effect of $H_2$ in atomically thin TMDC synthesis, its function in the salt assisted growth is not ubiquitous. To elucidate the role of $H_2$ in the monolayer formation, we first tested the sole effect of $H_2$. When we introduce $H_2$ to the molten $Na_2WO_4$, cubic crystals varying in colour from yellow to orange emerge. XPS and Raman measurements show that these crystals are sodium tungsten bronzes ($Na_xWO_3$, x<1) of various Na ratio (see Supplementary Information **Figure S6**)[27]. Accordingly, we determine that the temperature in which $H_2$ dosed during the TMDC synthesis is critical to prevent premature reduction of sodium tungstate to sodium tungsten bronze.

Curiously, when we introduce $H_2$ to $Na_2WO_4$ after 20 minutes of Se exposure, we don't observe $WSe_2$ formation. XPS and EDX analysis (**Figure 3a-b**) of the $Na_2WO_4$ liquid exposed to Se for 20 minutes without $H_2$, show that no Se dissolves in $Na_2WO_4$. Even when we add Se to the $NaCl:WO_3$ mixture, no Se is detected in the liquid droplets. We would like to note that when $H_2$ is introduced over hot Se vapour above 300 °C, $H_2$ reacts with Se to form $H_2Se$ gas[28]. $H_2Se$ formation during the synthesis can be imperative in $WSe_2$ growth.

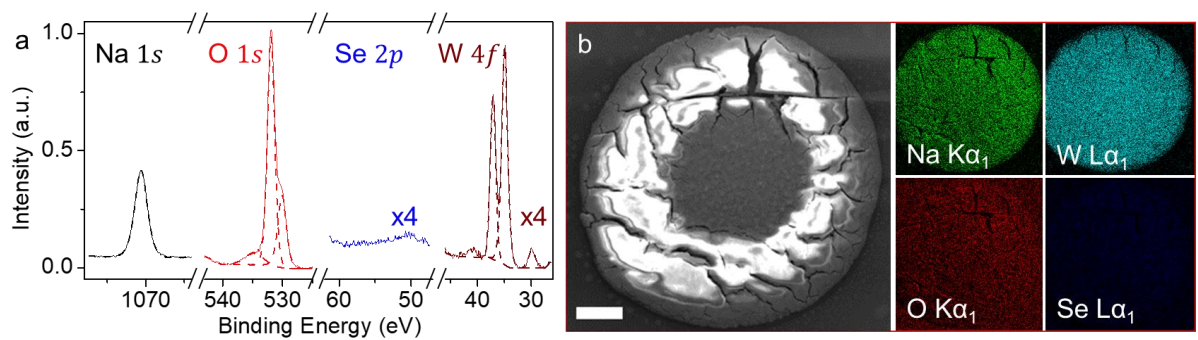

**Figure 3 | Effects of Se exposure on $Na_2WO_4$. a,** High resolution XPS spectra of $Na_2WO_4$ liquid droplet taken after 20 minutes of exposure to Se. Na $1s$, O $1s$, Se $2p$, W $4f$ spectra match exactly with $Na_2WO_4$ spectra and no Se $2p$ peaks exist in the spectrum. **b,** SEM image and EDX maps corresponding to the labelled elements are taken from the solidified $Na_2WO_4$ liquid droplet exposed to Se for 20 minutes. Se L$α_1$ map shows a faint background as the tail of W L$α_1$ peak extends through Se L$α_1$ energy. Scale bar is 20 μm.

We performed a series of controlled experiments to unravel any possible effect of $H_2Se$ on the $WSe_2$ formation. First, we evacuated the chamber and flushed it with Ar several times. Then, we filled the chamber with 5:1 ratio of Ar:$H_2$ till the atmospheric pressure and brought Se heater to 350 °C for Se vapour to react with $H_2$ in the CVD chamber to form $H_2Se$. We kept the substrate heater at 300 °C to minimize Se condensation on the substrate. After 15 minutes, we shut down the Se heater and ramped up the substrate heater to 750 °C. Unreacted Se vapour condenses on the cold chamber walls. As soon as the substrate heater went above 650 °C we started observing formation of $WSe_2$ monolayers from the forming $Na_2WO_4$ liquid (VLS) as well as at remote positions (VSS) on the substrate where no liquid can be found. As a controlled experiment the Se heater is heated to 350 °C as before, this time in the absence of $H_2$. After 15 minutes the Se heater is shut down and 5:1 Ar:$H_2$ is introduced in to the chamber and the substrate heater is heated to 750 °C. However, the controlled experiment didn't result in any $WSe_2$ crystal formation. This experiment shows that $H_2Se$ gas is required to synthesize

WSe$_2$ crystals both in VLS and VSS modes. Since selenium has a lower reactivity compared to sulphur[29,30], unlike sulphur containing TMDC synthesis, dependence of WSe$_2$ formation on H$_2$Se is not surprising.

All of the findings discussed above indicate that for the VLS mode synthesis, H$_2$Se and H$_2$ gasses react with the liquid intermediate compound Na$_2$WO$_4$ to form WSe$_2$ following the proposed reaction:

$$Na_2WO_{4\,(l)} + 2\,H_2Se_{\,(g)} + H_{2\,(g)} \rightarrow WSe_{2\,(s)} + 3\,H_2O_{\,(g)} + Na_2O_{\,(s)} \quad (2)$$

Here, H$_2$ is also needed as a reducing agent to reduce W$^{6+}$ in Na$_2$WO$_4$ to W$^{4+}$ in WSe$_2$. For the VSS mode, the gaseous intermediate product WO$_2$Cl$_2$ reacts with H$_2$Se and H$_2$ to form WSe$_2$ crystals. The chemical reaction of the adsorbed molecules that produce WSe$_2$ can be written as:

$$WO_2Cl_{2\,(ads)} + 2\,H_2Se_{\,(ads)} + H_{2\,(g)} \rightarrow WSe_{2\,(s)} + 2\,H_2O_{\,(g)} + 2\,HCl_{\,(g)} \quad (3)$$

We would like to note that the crystals grow at a much faster rate in the VSS mode compared to the VLS mode. This observation is consistent with the previous reports[9]. These two proposed reactions are critical to understand the effect of each growth parameter that leads to the WSe$_2$ synthesis.

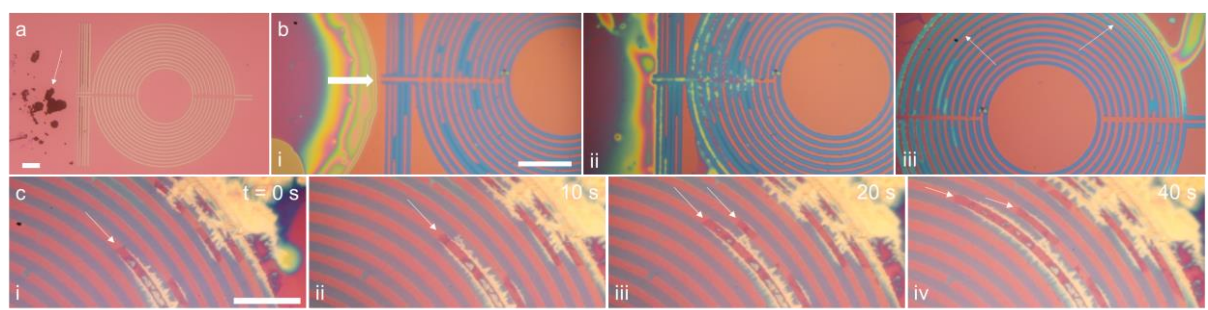

**Figure 4 | Control of the intermediate compound for patterned TMDC synthesis. a,** Optical micrograph of HfO$_2$ channels patterned on an oxidised silicon chip. The white arrow indicates 1:2 weight ratio mixture of NaCl:WO$_3$ grain. **b,** A series of optical micrographs (**i-iii**) taken during the liquid Na$_2$WO$_4$ spreading through the channels at 750 °C. The white arrow in **i** indicates the direction liquid enters the channel and the arrows in **iii** show a thin liquid residue left on the channel walls. **c,** A series of images (**i-iv**) taken several seconds apart shows how the liquid precursor turns into WSe$_2$ by following the channels. Scale bars are 100 µm in **a, b,** and 50 µm in **c**.

We are now in a position to control the liquid intermediate compound to synthesize TMDCs in pre-determined patterns[31–35]. To illustrate how our custom-made CVD chamber can be employed to study patterned crystal growth via directing sodium tungstate liquid, we fabricated HfO$_2$ channels on oxidised Si chips using standard optical and electron beam lithography. **Figure 4a** shows a patterned substrate with channels formed by 10 µm wide, 23.5 nm tall HfO$_2$ walls. Fine NaCl:WO$_3$ mixture of 1:2 weight ratio is placed at one end of the channels. The reaction between the salt and the metal oxide precursor produces Na$_2$WO$_4$ droplets of various sizes on the substrate surface, whose viscosity depends on the unreacted WO$_3$ concentration in the droplet. As the temperature goes above 700 °C, Na$_2$WO$_4$ spreads through the channel upon contact with the channel ends due to the capillary action (**Figure 4b**). After the introduction of H$_2$ and Se, particularly the liquid that crawls along the channel walls forms the WSe$_2$ crystals. **Figure 4a** shows formation of crystals in circular channels (see Supplementary Movie 3 for the patterned crystal synthesis). Although the monolayer formations presented here have multi-layer patches, it would be possible to control the layer thickness by adjusting the channel wall dimensions.

The methodology we presented here can be applied to any other TMDC that can be synthesized in a CVD chamber. Indeed, we tested the chamber to synthesize other selenium based TMDCs such as $MoSe_2$ [26,36–39] and $VSe_2$ [5,9,40,41] (see Supplementary Information). In **Figure 5a**, a series of real time optical micrographs taken during the synthesis shows grains of $NaCl:MoO_3$ mixture reaction to form the liquid intermediate compound. We found out that although the growth conditions for salt assisted synthesis of $MoSe_2$ are different than that of $WSe_2$, mechanisms are similar. **Figure 5b** shows a series of optical images taken during the VLS formation of $MoSe_2$ monolayers.

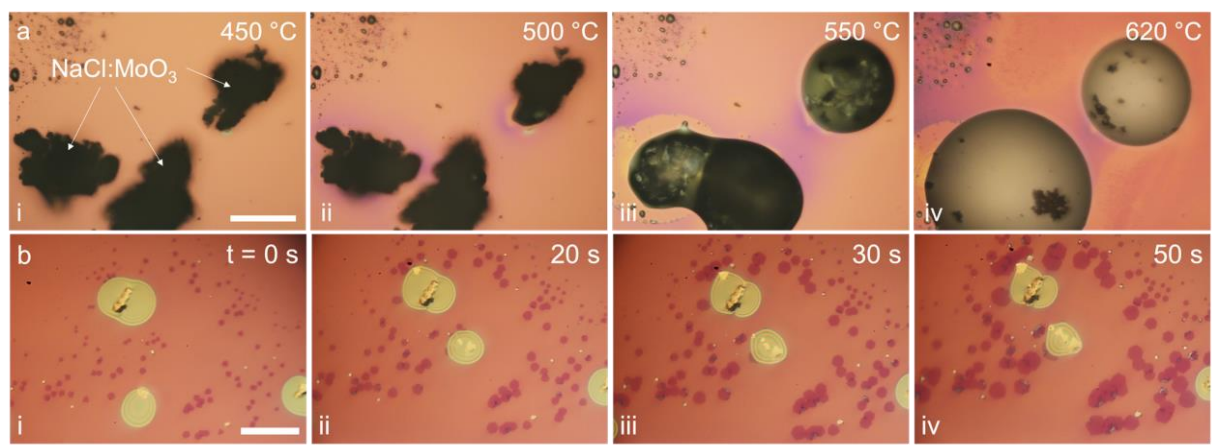

**Figure 5 | Real time optical micrographs of MoSe$_2$ synthesis. a,** A series of real time optical micrographs (**i-iv**) taken at various substrate heater temperatures show that 1:2 weight ratio mixture of $NaCl:MoO_3$ grains placed on the substrate reacts to form liquid droplets. Scale bar is 100 µm. **b,** Optical micrographs (**i-iv**) taken after introducing Se and $H_2$ at 650 °C substrate temperature show how crystals rapidly form by consuming the liquid precursor. Within t = 50 seconds from the beginning of the observation, crystals of size almost as large as 50 µm form. Scale bar represent 50 µm.

In summary, we demonstrated that our custom-made CVD chamber can be used to monitor and control the synthesis routes in real time. We showed that both VSS and VLS routes can be employed to grow selenium based TMDC crystals. In VSS route, $H_2Se$ and $H_2$ reacts with the metal-oxychloride to form the monolayers while in VLS mode a liquid composed of the alkali metal, transition metal and oxygen reacts with $H_2Se$ and $H_2$ to form the monolayers. We realize that by timing when $H_2$ and Se is introduced, it is possible to control the growth route. Such a degree of control over the synthesis route, allowed us to guide the liquid intermediate product to synthesize $WSe_2$ along pre-patterned channels. We would like to emphasize that although the reaction intermediate products and the reaction routes may differ among various TMDCs, the methods we present here will still be applicable. Furthermore, the CVD chamber we reported in this article can be modified to investigate the synthesis mechanisms of other materials that utilizes a CVD chamber for the growth. Optical observations can be accompanied by other spectroscopic measurement techniques that can be incorporated on to the chamber to provide real time spectroscopic information about the intermediate phases.

**Methods**

CVD chamber is machined out of a 6061-aluminium alloy. There are three body pieces that form the chamber: bottom plate, reactor and lid. The bottom plate houses the water circulation channels and the cooling water makes direct contact with the reactor when the two pieces are bolted together. The lid allows an easy access to the reactor chamber. There are water jackets within the lid as well. The chamber body temperature is maintained at or near room temperature. A closed-cycle chiller supplies water to the cooling channels embedded within the chamber body and the lid. Outside the chamber, temperature of hottest regions remains below 50 °C. The optical port on the lid is a 0.5 mm thick sapphire disk and it is directly located above the substrate heater. Electrical connections for temperature control are made via hermetically sealed feedthroughs. There are two separate gas inlets, one for Ar and the other for $H_2$ and a single exhaust at the opposite end of the chamber. Gases are dosed through mass flow controllers. The chamber can reach down to $10^{-3}$ mBar using an oil rotary vane pump. Heater temperatures and gas flow rates are controlled via a LabView software. Detailed pictures of the chamber are given in the Supplementary Information.

**Acknowledgements**

This work is supported by Turkish Scientific and Technological Research Council under grant no: 116M226. We would like to thank Engin Can Sürmeli and Ali Sheraz for discussions, Fatih Yaman for his help with the LabView programming, Talha Masood Khan for providing masks for optical lithography and Bülend Ortaç for providing $MoO_3$ precursor.

**Author Contributions**

TSK conceived the experiments. HRR and TSK designed the CVD chamber and assembled it. HRR and TSK performed the experiments. NM and OÇ helped characterizing the samples. All the authors discussed the results and contributed to the writing of the manuscript.

# Supplementary Information: Real time optical observation and control of atomically thin metal chalcogenide synthesis

Hamid Reza Rasouli[1], Naveed Mehmood[1], Onur Çakıroğlu[2], T. Serkan Kasırga[1,2]*

[1] UNAM – Institute of Materials Science and Nanotechnology, Bilkent University, Ankara 06800, Turkey

[2] Department of Physics, Bilkent University, Ankara 06800, Turkey

* Corresponding Author: kasirga@unam.bilkent.edu.tr

**Table of Contents**

**Supplementary Text**

**1. CVD Chamber**

The chamber is composed of 3 main parts: (1) cooling base, (2) reactor, (3) lid. Cooling base holds the main cooling channels for the reactor. Reactor is the housing that has the hermetically sealed electrical feedthroughs and the gas inlet and outlets. The lid is also composed of 3 main parts. The lid body, sapphire window and sapphire window holder is assembled in a vacuum tight way. The cooling channels within the lid helps to keep the lid cool. All the body parts are machined out of 6061 Aluminium alloy as it offers 10 times higher thermal conductivity compared to 304 stainless steel. Images of the setup is given in **Figure S1**.

### 1.1. Alumina Heaters

We cast alumina heaters, using a mixture of 85% $Al_2O_3$, 5% kaolin and 10% water mixture with Kanthal A1 22 Ga wire coils placed within the mould. Then, we dry the green body at 100 °C to evaporate the excess water and mixture to set. Finally, the green body is sintered at 1400 °C for three hours. This forms a solid body, yet as the sintering temperature is kept low to prevent deterioration of Kanthal wires it is not as strong as alumina produced by other methods.

After the heaters are sintered, we place thermocouple 1 mm deep into the surface and by using a hand-held infrared thermometer we check the accuracy of the reading by the thermocouple. If the reading is accurate within ± 10 °C, we place the heater in the chamber. For the substrate heaters we perform one final calibration using NaCl. A few grains of NaCl is placed on an oxidised chip and its melting point is observed on the substrate heater. If it agrees ± 20 °C with the melting point of NaCl, we start using the heater. Typically, the heaters last for many growth cycles. Indeed, most of the studies we reported in this paper has been performed using the same set of heaters. Care must be taken especially during cooldown. Sudden changes in the heater current result in expansion and contraction of the Kanthal wire, which weakens the heater over time.

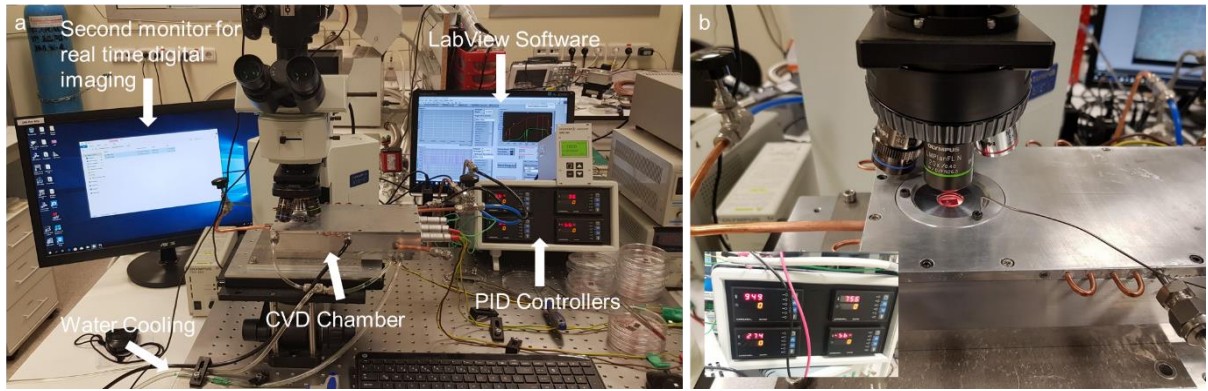

**Figure S1 | Pictures of the custom-made CVD setup. a,** Picture shows the layout of the overall setup. Some essential components are labelled on the picture. **b,** Close-up view of the chamber running at high temperature. Inset picture shows the PID controller assembly that indicates the temperatures of the substrate heater (top left), heater 1 (top right) and heater 2 (bottom left) during a synthesis.

### 1.2. Auxiliary components

Whole CVD chamber is orchestrated via a LabView program that controls the mass flow controllers and the heater temperatures. Also, we use a 700 W/h chiller to keep the chamber walls cold. Typically, we keep the coolant temperature around 10 °C. The CVD chamber is assembled on an Olympus BX 51WI fixed stage microscope. The focusing is performed via moving the nosepiece rather than the stage. For the X-Y motion, the whole chamber moves thanks to the flexible piping and cabling. The microscope is equipped with a 5x, a 20x and an ultra-long working distance 40x objective with 20x multiplication at the eyepiece. A Canon DSLR camera is connected to the microscope for capturing HD video and high-resolution pictures.

### 1.3. Thermal simulations

We modelled the heater configuration in an Ar atmosphere by using a finite element analysis software (COMSOL). Although we simulated by gas flow as well, since the flow rate is low it makes no difference on the temperature profile. For the analysis we kept the top lid at 50 °C and the heater at 800 °C. Thermal analysis shows that when the surface of the heater is at 800 °C, temperature within the first mm is above 600 °C. Results of the analysis are given in **Figure S2**. The thermal gradient towards the lid creates a convection current. This convection carries some of the vaporized precursors up to the lid and the observation window. For many materials we synthesized, this causes no problem while for some that has high vapor pressure, condensation of the evaporated matter on to the observation window hinders the vision.

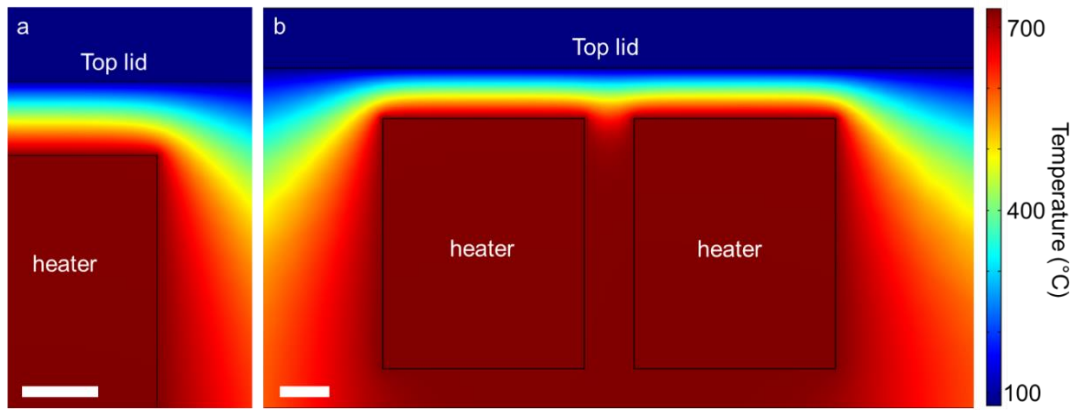

**Figure S2 | Cross-sectional view of the thermal distribution within the CVD chamber. a,** Single heater configuration and **b,** double heater configuration. Scale bar represents 3 mm.

## 2. Characterization of a typical WSe$_2$ crystal

**Figure S3a** shows an optical picture of a typical crystal synthesized in our custom CVD chamber. Raman and photoluminescence (PL) maps (**Figure S3b-c**) show that the crystal is uniform, and the PL spectrum is similar to that has been reported previously[1]. We also performed atomic force microscopy on our crystals to measure their thicknesses. AFM measurements show that (**Figure S4**) the monolayer WSe$_2$ has a thickness of 0.75 nm. It is interesting to note that the surface of the crystal is covered with small dots of unknown composition.

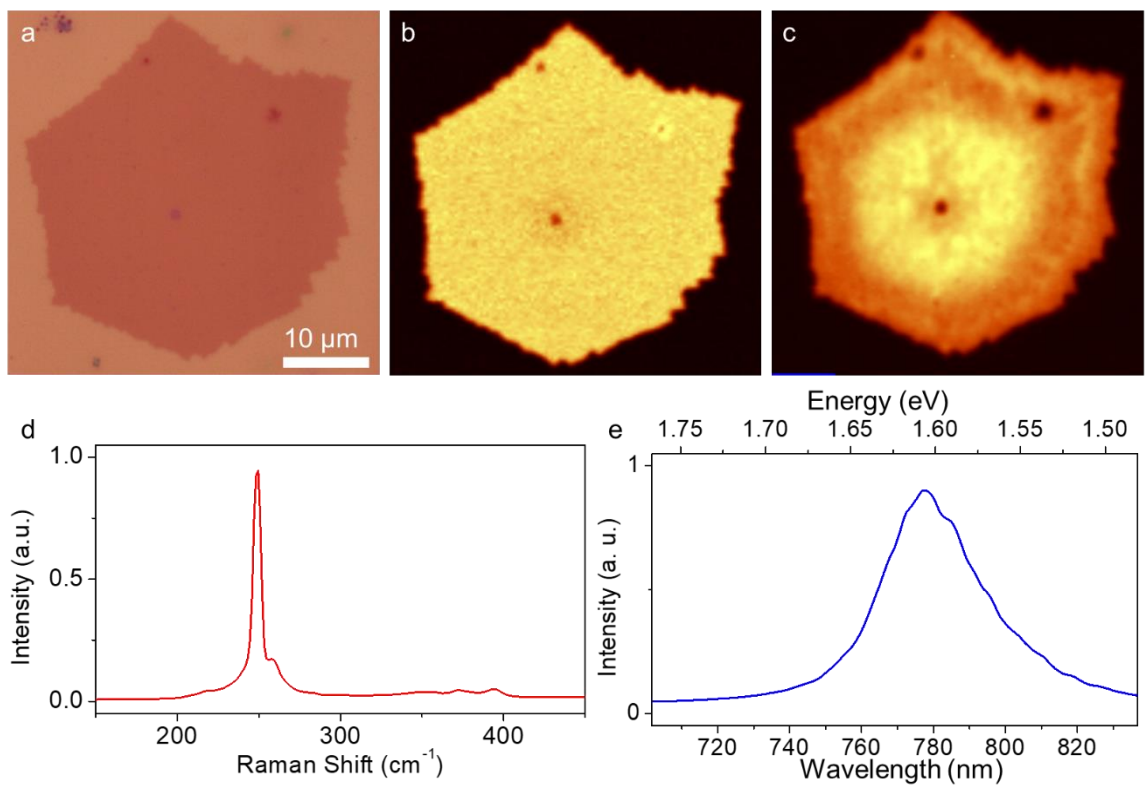

**Figure S3 | Characterization of a typical WSe$_2$ monolayer. a,** Optical microscope image of a typical crystal is given. **b,** Raman and **c,** PL maps show the high quality of the crystals. **d,** Raman spectrum taken from a point on the crystal shows the typical Raman features of the monolayer WSe$_2$. **e,** PL spectrum taken from a point on the crystal is similar to the previously reported CVD grown monolayers of WSe$_2$.

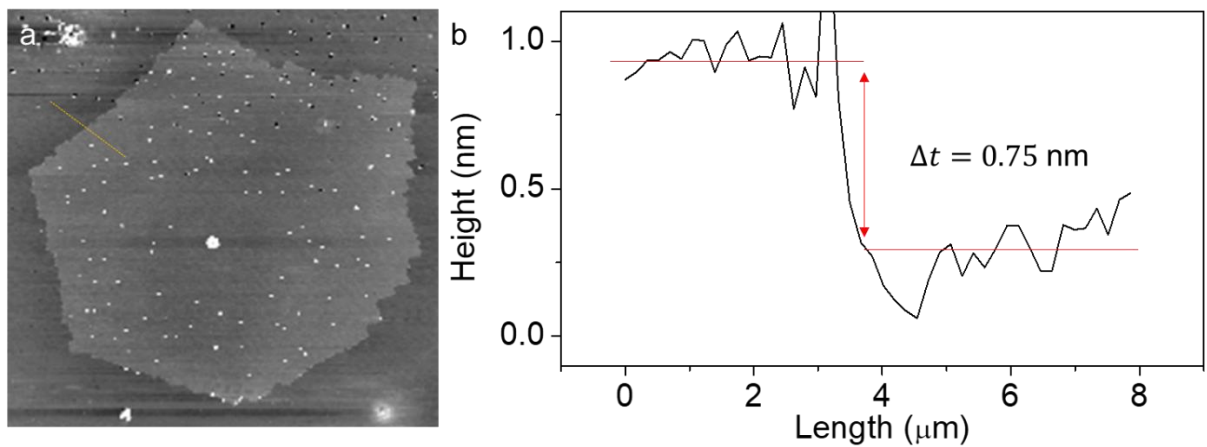

**Figure S4 | AFM topography of a monolayer WSe$_2$. a,** AFM height map and **b**, the height trace taken along the yellow dashed line in **a** shows that the apparent thickness of the crystal is 0.75 nm.

3. **Early stage of NaCl – WO$_3$ reaction**

**Figure 5S** shows EDX analysis on a WO$_3$ particle at an early stage of the synthesis. EDX maps in **Figure 5S b-g** shows that Na and Cl coexist in certain locations on WO$_3$ grain.

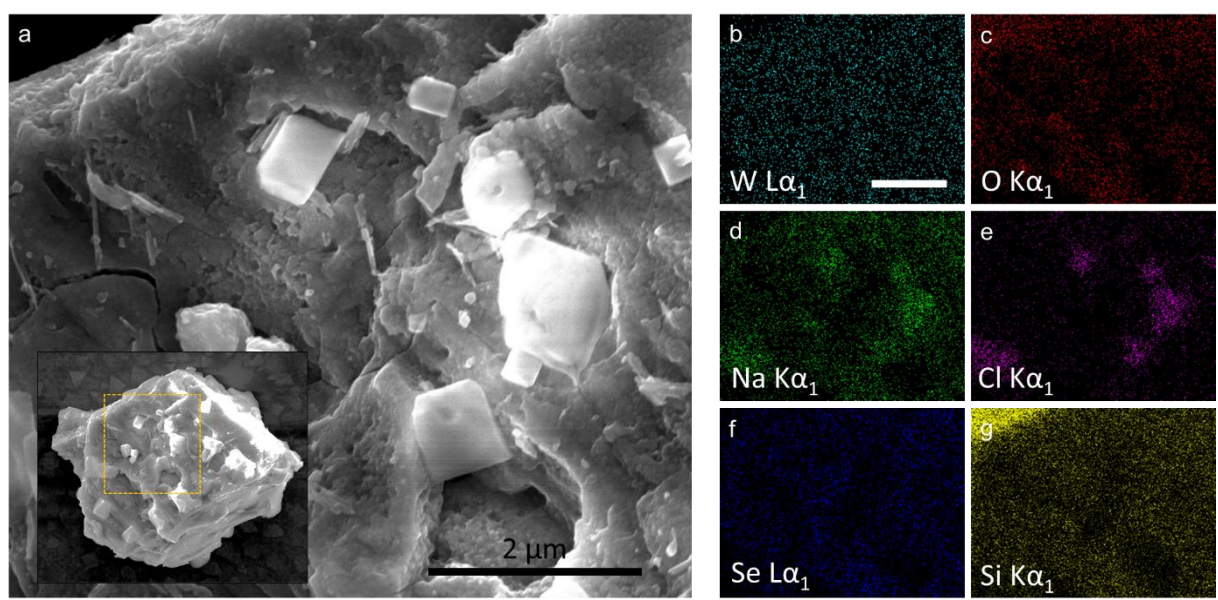

**Figure S5 | EDX of early stage NaCl – WO$_3$ reaction. a,** SEM image of a WO$_3$ grain shows crystal formations on the surface **b-g,** EDX maps taken from the same region shows that there are regions where NaCl particles recrystallized on the WO$_3$ surface. Comparing **d** and **e** reveals that not all the Na on the surface is in NaCl form. **f**, Although no Se is introduced, as the Se L$\alpha_1$ peak coincides with the tail of the W L$\alpha_1$ peak, we observe an erroneous distribution of Se all over the material.

4. **Formation of Sodium Tungsten bronze with H$_2$**

As mentioned in the main text, when we introduce hydrogen gas to Na$_2$WO$_4$, crystals of various colours ranging from yellow to grey form as shown in **Figure S6a-b**. EDX (**Figure S6c-g**) and XPS (**Figure S6h**) measurements show that these crystals are sodium tungsten bronzes which has a form of Na$_x$WO$_3$, $x < 1$. Varying colours correspond to different $x$.[2] W $4f$ binding energies are consistent with the sodium tungsten bronze[3] and Raman spectrum (**Figure S6i**) taken from the crystals are also consistent with the Raman spectra of alkali tungsten bronzes[4].

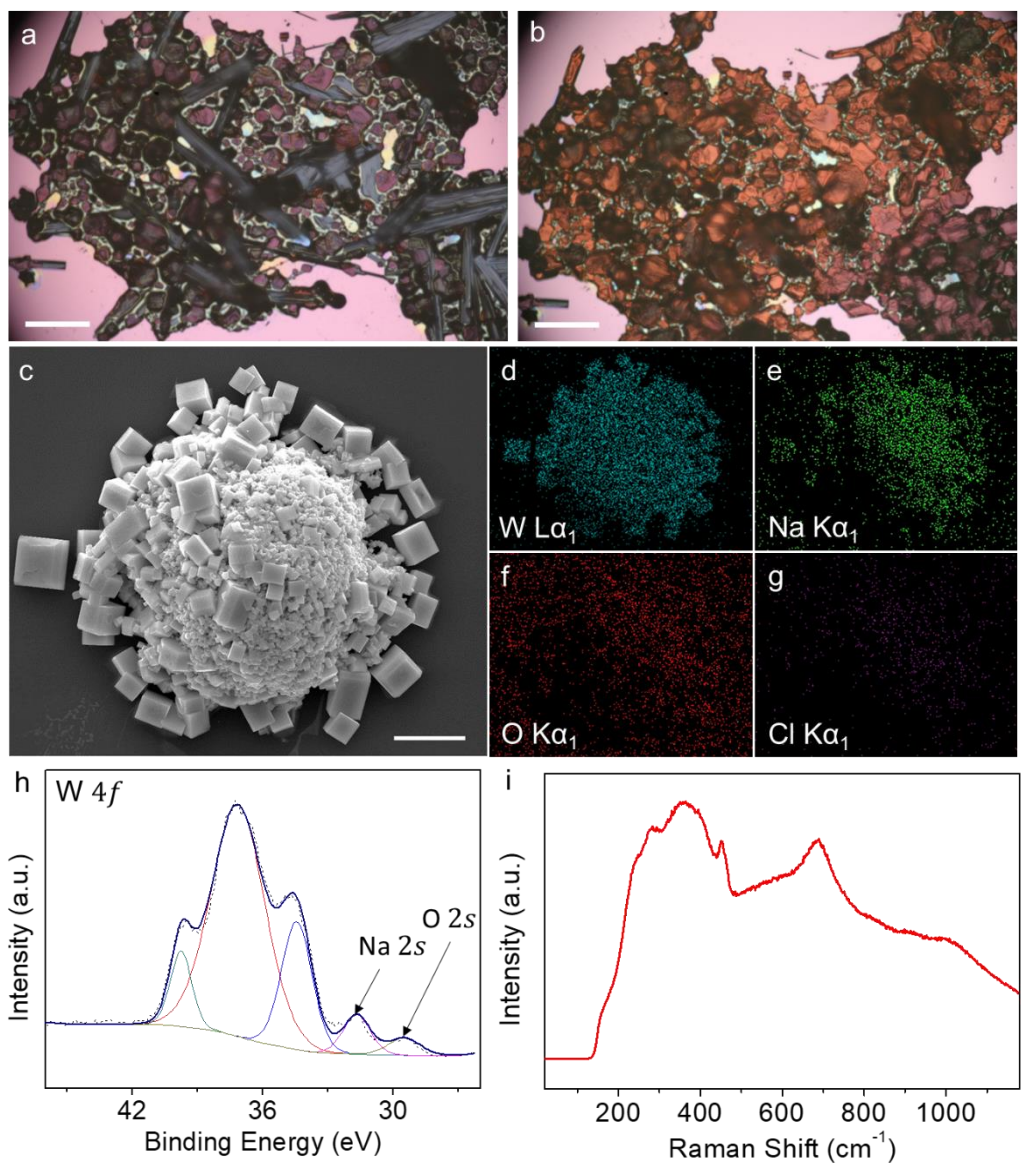

**Figure S6 | Sodium Tungsten bronze formation. a, b,** Optical micrographs show crystals of various colours after exposing $Na_2WO_4$ liquid droplet to $H_2$. The colours we observe in these formations are consistent with the reported colours for $Na_xWO_3$. Scale bars are 100 µm. **c.** SEM micrograph and **d-g** corresponding EDX maps. **h,** XPS survey of W $4f$ binding energy taken from sodium tungsten bronze formation indicates that W is in +4 oxidation state. **i,** Raman spectrum of sodium bronze.

## 5. Characterization of a typical MoSe₂ crystal

**Figure S7a** shows an optical picture of a typical MoSe$_2$ crystal synthesized in our custom CVD chamber. AFM measurements given in **Figure S7b** shows a monolayer thickness of 0.72 nm. Raman and photoluminescence (PL) maps (**Figure S7c-d**) show that the crystal is uniform, and the PL spectrum is similar to that has been reported previously [5,6].

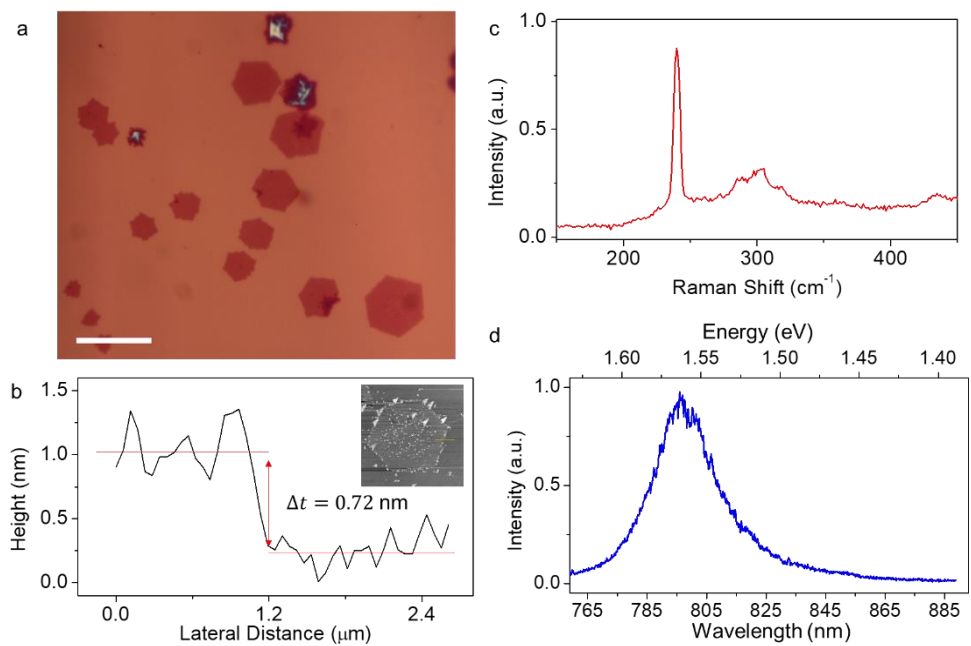

**Figure S7 | Characterization of a typical MoSe$_2$ crystal. a,** Optical micrographs of hexagonal MoSe$_2$ monolayers on SiO$_2$/Si. **b,** AFM height trace taken along the dashed line in the inset topography map shows that the average height of the crystal is 0.72 nm. **c**, Raman spectrum and **d**, photoluminescence spectrum are consistent with the CVD growth MoSe$_2$ monolayers.

## 6. Characterization of a typical VSe$_2$ crystal

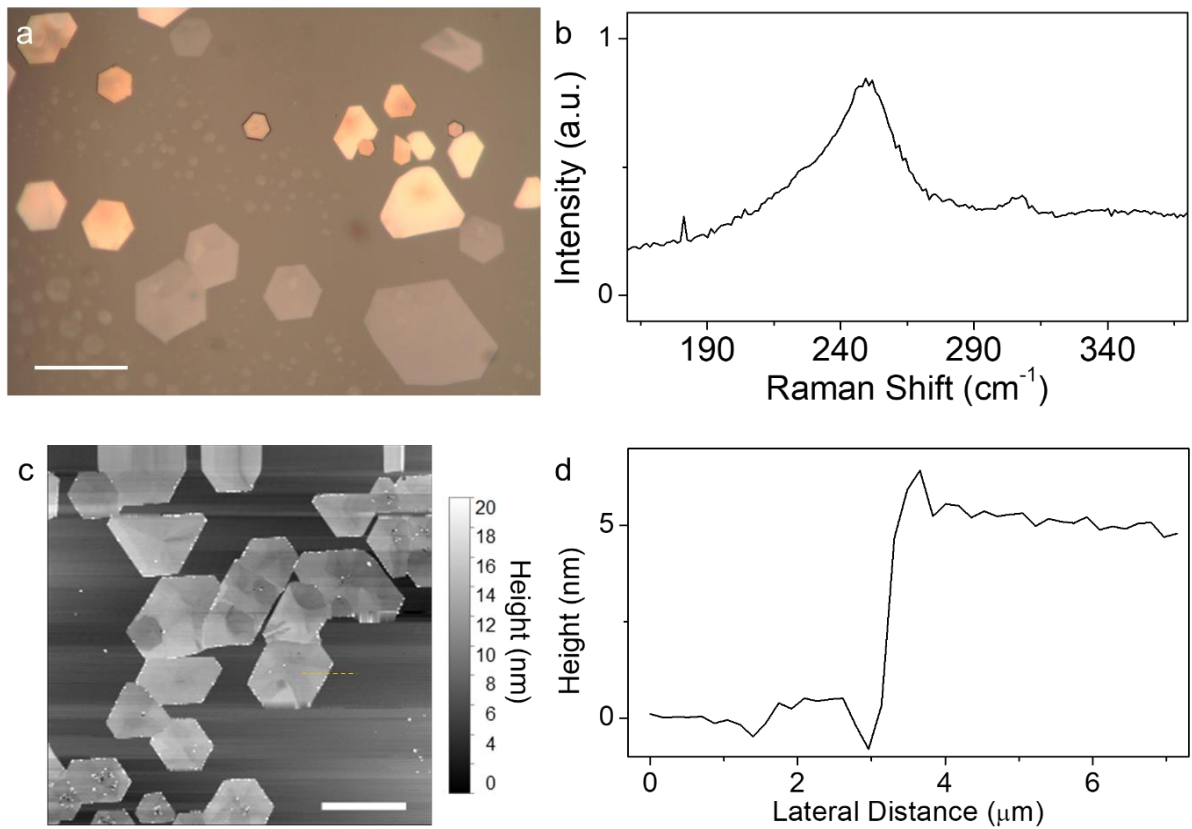

**Figure S8 | Characterization of VSe$_2$ crystals on c-cut sapphire substrate. a**, Optical micrograph of the crystals on sapphire surface. Scale bar is 20 μm. **b,** Raman spectrum of a 5

nm thick crystal. This spectrum is consistent with a recent report.[7] **c,** AFM height map and **d,** AFM height trace taken along the yellow dashed line in **c**. The thinnest crystal we obtained so far has been 5 nm.

---

[1] Liu, B. et al. ACS Nano, 10.1021/acsnano.5b01301
[2] Xue, Y., Zhang, Y et al. Phys. Rev. B 10.1103/PhysRevB.79.205113
[3] Shanks, H. R. 1977 Phys. Rev. B 10.1103/PhysRevB.16.697
[4] Shahidur Rahman, 2015 Thesis
[5] Shaw, J. et al. Nano Research, https://doi.org/10.1007/s122
[6] Chang, Y.H. et al. ACS Nano 10.1021/nn503287m
[7] Zhou, J. *et al.* A library of atomically thin metal chalcogenides. *Nature* **556,** 355–359 (2018).